# Instabilities, breathers and rogue waves in optics

**John M. Dudley[1], Frédéric Dias[2], Miro Erkintalo[3], Goëry Genty[4]**

1. Institut FEMTO-ST, UMR 6174 CNRS-Université de Franche-Comté, Besançon, France

2. School of Mathematical Sciences, University College Dublin, Belfield, Dublin 4, Ireland

3. Department of Physics, University of Auckland, Auckland, New Zealand

4. Department of Physics, Tampere University of Technology, Tampere, Finland

**Optical rogue waves are rare yet extreme fluctuations in the value of an optical field. The terminology was first used in the context of an analogy between pulse propagation in optical fibre and wave group propagation on deep water, but has since been generalized to describe many other processes in optics. This paper provides an overview of this field, concentrating primarily on propagation in optical fibre systems that exhibit nonlinear breather and soliton dynamics, but also discussing other optical systems where extreme events have been reported. Although statistical features such as long-tailed probability distributions are often considered the defining feature of rogue waves, we emphasise the underlying physical processes that drive the appearance of extreme optical structures.**

Many physical systems exhibit behaviour associated with the emergence of high amplitude events that occur with low probability but that have dramatic impact. Perhaps the most celebrated examples of such processes are the giant oceanic "rogue waves" that emerge unexpectedly from the sea with great destructive power [1]. There is general agreement that the emergence of giant waves involves physics different from that generating the usual population of ocean waves, but equally there is a consensus that one unique causative mechanism is unlikely. Indeed, oceanic rogue waves have been shown to arise in many different ways: from linear effects such as directional focusing or random superposition of independent wave trains, to nonlinear effects associated with the growth of surface noise to form localized wave structures [1,2].

The analogous physics of nonlinear wave propagation in optics and in hydrodynamics has been known for decades, and the focusing nonlinear Schrödinger equation (NLSE) applies to both systems in certain limits (Box 1). The description of instabilities in optics as "rogue waves" is recent, however, first used in 2007 when shot-to-shot measurements of fibre supercontinuum (SC) spectra by Solli *et al.* yielded long-tailed histograms for intensity fluctuations at long wavelengths [3]. An analogy between this optical instability and oceanic rogue waves was suggested for two reasons. Firstly, highly skewed distributions are often considered to define extreme processes, since they predict that high amplitude events far from the median are still observed with non-negligible probability [4]. And secondly, the particular regime of SC generation being studied developed from modulation instability (MI), a nonlinear process associated with exponential amplification of noise that had previously been proposed as an ocean rogue wave generating mechanism [2].

These pioneering results enabled for the first time a quantitative analysis of the fluctuations at the spectral edge of a broadband supercontinuum, and motivated many subsequent studies into how large amplitude structures could emerge in optical systems. These studies attracted broad interest and have essentially opened up a new field of "optical rogue wave physics". Although most research since has focused on propagation in optical fibres and in particular in regimes analogous to hydrodynamics, the terminology "optical rogue wave" has now been generalized to describe other noisy processes in optics with long tailed probability distributions, irrespective of whether they are observed in systems with a possible oceanic analogy. Moreover, particular analytic solutions of the NLSE describing solitons on a finite background or "breathers" are now also widely referred to as "rogue wave" solutions, even when studied outside a statistical context for mathematical interest, or when generated experimentally from controlled initial conditions. These wider definitions have become well-established, but can unfortunately lead to difficulty for the non-specialist.

Our aim here is to remove any possible confusion by presenting a synthetic review of the field, but we do so not in terms of its chronological development which has been discussed elsewhere [5,6]. Rather, we organise our presentation by classifying rogue waves in terms of their generating physical mechanisms. We begin by discussing rogue waves in the regime of NLSE fibre propagation where MI and breather evolution dominate the dynamics, and we then discuss how the effect of perturbations to the NLSE can lead to the emergence of background-free solitons. This provides a natural lead-in to discuss the physics and measurement techniques of rogue solitons in fibre supercontinuum generation. We then describe techniques used to control rogue waves dynamics in fibre systems, and this is followed by a survey of results in other systems: lasers and amplifiers where dissipative effects are central to the dynamics, and spatial systems where both nonlinear and linear dynamics can play a role.

**ROGUE WAVES AND STATISTICS**

Before considering specific examples of optical rogue waves, we first briefly review how rogue wave events are manifested in the statistics of the particular system under study.

In optics, statistical properties are defining features of light sources. For example, the random intensity fluctuations of polarized thermal light follow an exponential probability distribution, and the intensity fluctuations of a laser above threshold follow a Gaussian probability distribution [7]. It was the experimental observation of "L-shaped" long tailed distributions in Ref. [3] that linked for the first time nonlinear optics with the wider theory of extreme events. In a sense, seeing long-tailed distributions in optics is not surprising, since it is well-known that a nonlinear transfer function will modify the probability distribution of an input signal. Indeed, an exponential probability distribution in intensity is transformed under exponential gain to a power law Pareto-distribution. There are other cases, however, where a functional nonlinear transformation of an input field cannot be identified, and the emergence of high amplitude events arises from more complex dynamics. Optical rogue waves and long-tailed statistics have been observed in systems exhibiting both types of behaviour.

Rogue waves in optics have been identified in different ways. One approach uses the idea from probability theory that associates rogue events with particular extreme-value probability distributions, and such functions have provided good fits to the tails of histograms of optical intensity fluctuations in several studies [8-10]. Another approach has adapted the oceanographic definition, where rogue waves are those with trough-to-crest height $H_{RW}$ satisfying $H_{RW} \geq 2\ H_{1/3}$. Here $H_{1/3}$ is the significant wave height, the mean height of the highest third of waves [11]. In optics, however, the accessible data is not the field amplitude but rather the intensity, and indeed such data can take a variety of forms: an intensity time series; the levels of a two-dimensional camera image; the space-time intensity evolution of an optical field. From this data, the intensity peaks are analysed statistically to compute a histogram, and the oceanographic definition is modified to: $I_{RW} \geq 2\ I_{1/3}$ where the "significant intensity" $I_{1/3}$ is the mean intensity of the highest third of events. Although somewhat arbitrary, this definition has been applied in several studies [12-15].

From a general perspective, what is of most interest is to consider whether events in the distribution tail for a particular system arise from different physics compared to the events closer to the distribution median. This is also relevant in a practical context, since the ability to identify the conditions causing extreme events is key to their prediction and control.

**MODULATION INSTABILITY AND BREATHER DYNAMICS**

Modulation instability is a fundamental property of many nonlinear dispersive systems, associated with the growth of periodic perturbations on a continuous wave (CW) background [16]. In the initial evolution, the spectral sidebands associated with the instability experience exponential amplification at the expense of the pump, but the subsequent dynamics are more complex and display cyclic energy exchange between multiple spectral modes. In optics, MI seeded from noise results in a series of high-contrast peaks of random intensity [17,18], and it is these localised peaks that have been compared with similar structures seen in studies of ocean rogue waves [2,19,20]. Significantly, the MI growth and decay dynamics in the NLSE have exact solutions in the form of various types of "breather" or "soliton on finite background" (SFB) [21], and this fact has motivated much research to gain *analytic* insight into conditions favouring rogue wave emergence.

SFBs constitute a class of NLSE solutions whose real and imaginary parts are linearly related, and they include Akhmediev Breathers (ABs), Kuznetsov-Ma (KM) solitons, the Peregrine soliton (PS), and even more generally the bi-periodic solutions of the NLSE described by Jacobi elliptic functions [21-23]. Many of these solutions have been labelled as "rogue waves" [24] although this interpretation must be made with great care. As we will see below, the statistical criterion for rogue waves in an MI field seeded by noise is generally only satisfied by particular higher-order SFB solutions (sometimes referred to as "multi-rogue waves" or "higher-order rogue waves") [25-28].

Before discussing noise-seeded MI in detail, we first describe these SFBs and breather solutions by referring to the dimensionless NLSE:

$$i\frac{\partial \psi}{\partial \xi} + \frac{1}{2}\frac{\partial^2 \psi}{\partial \tau^2} + |\psi|^2 \psi = 0. \tag{1}$$

The envelope $\psi(\xi,\tau)$ is a function of $\xi$ (propagation distance) and $\tau$ (co-moving time). Equation (1) can be related to the dimensional fibre-optic NLSE in Fig. B1 by defining timescale $T_0 = (|\beta_2| L_{NL})^{1/2}$ and nonlinear length $L_{NL} = (\gamma P_0)^{-1}$, where $P_0$ is optical background power in W. The dimensional field $A(z,T)$ [W$^{1/2}$] is $A = \sqrt{P_0}\psi$; dimensional time $T$ [s] is $T = \tau T_0$ and dimensional distance $z$ [m] is $z = \xi L_{NL}$. Analytic solutions of MI dynamics have been obtained by several authors [20-23,29-31], with the relevance to optics first pointed out in Ref. [30]. The particular solution that describes MI growth and decay is [21,30]:

$$\psi(\xi,\tau) = e^{i\xi}\left[1 + \frac{2(1-2a)\cosh(b\xi) + ib\sinh(b\xi)}{\sqrt{2a}\cos(\omega\tau) - \cosh(b\xi)}\right] \tag{2}$$

The solution's properties are determined by one positive parameter $a$ ($a \neq 1/2$) through arguments $b = [8a(1-2a)]^{1/2}$ and $\omega = 2(1-2a)^{1/2}$. Over the range $0 < a < \frac{1}{2}$ the solution is the Akhmediev breather (AB) where we observe evolution from the trivial plane wave ($a = 0$) to a train of localised pulses with temporal period $\Delta\tau = \pi/(1-2a)^{\frac{1}{2}}$ [30]. The AB solution provides MI's analytic framework, with the real parameter $\omega$ corresponding to the modulation frequency, and the real parameter $b$ giving the parametric gain coefficient [32].

Figure 1(a) plots solutions for different $a$ as indicated, and it is the spatial and temporal localization properties of these solutions which have led to their association with rogue waves. The MI instability growth rate is maximal at $a = \frac{1}{4}$, but increasing $a$ actually leads to stronger localization in both dimensions until the limit $a \rightarrow \frac{1}{2}$ which gives the Peregrine Soliton [23]. The PS, given by: $\psi(\xi,\tau) = [1 - 4(1+2i\xi)/(1 + 4\tau^2 + 4\xi^2)] e^{i\xi}$, corresponds to a single pulse with localization in time ($\tau$) as well as along the propagation direction ($\xi$) as shown, and it has maximum intensity amongst the AB solutions of $|\psi_{PS}|^2 = 9$.

When $a > ½$, the parameters $\omega$ and $b$ become imaginary, and the solution exhibits localisation in the temporal dimension $\tau$ but periodicity along the propagation direction $\xi$. This is the Kuznetsov-Ma soliton [22,29] which is shown in Fig. 1(a) for $a = 0.7$. In this regard, we note that it was the KM result in Ref. [22] which was the first mathematical SFB solution of the NLSE reported.

Each of the solutions described by Eq. (2) is a special case of a more general family of solutions that exhibit periodicity in *both* transverse time $\tau$ and longitudinal propagation direction $\xi$ [21]. More complex *higher-order* solutions also exist with even stronger localisation and higher intensities than the PS [21], such as higher-order rational solitons [25-28,33] and AB collisions [34]. Figure 1(a) shows examples of both types of solutions [25,35,36].

The analytical results above have been used to design experiments with controlled initial conditions to excite particular SFB dynamics in fibre optics. The use of optical fibres provides an especially convenient experimental platform as the dispersion and nonlinearity parameters can be conveniently matched to available optical sources to yield a propagation regime where the NLSE is a valid model for the dynamics. Experiments typically inject a multi-frequency field into a nonlinear fibre, similar to the method developed for coherent pulse train generation in telecommunications [37,38]. Figure 1(b-d) shows a selection of results obtained. The first experiments in 2010 used frequency-resolved optical gating to show localisation in the PS regime [39], with measured intensity and phase agreeing well with numerical and analytical predictions [Fig. 1(b)]. Later experiments examined the growth and decay of spectral amplitudes during AB evolution in more detail [40]. As pointed out by Van Simaeys *et al.*, these reversible dynamics of MI is a clear manifestation of Fermi-Pasta-Ulam recurrence in optics [41].

Experiments exciting KM-like evolution along the propagation direction have also been realised, and Fig. 1(c) shows results illustrating the growth and decay of the peak temporal intensity [42]. In another experiment, spectral shaping of an optical frequency comb synthesised initial conditions to excite the collision of two ABs [35]. Figure 1(d) shows an example of the results obtained, comparing the measured collision profile at the fibre output with numerical simulation. These results are significant in highlighting how collisions can yield significantly larger intensities than the elementary ABs alone, even exceeding the PS limit. Note that excited AB collision dynamics can be considered as a particular case of higher-order MI, where the simultaneous excitation of multiple instability modes within the MI gain bandwidth gives rise to the nonlinear superposition of ABs [21,43-45].

Note that the use of a modulated input field in these experiments means that the initial conditions correspond to a truncated Taylor-series expansion of the analytic AB or PS far from the point of maximum localization, or an approximation to the ideal KM profile at its point of minimal intensity in its evolution cycle. The use of non-ideal initial conditions induces differences compared to the ideal dynamics (e.g. the occurrence of multiple Fermi-Pasta-Ulam recurrence cycles for an AB), but simulations have shown that the spatio-temporal localisation and the field properties at point of highest intensity remain very well described by the corresponding analytic SFB solution [46].

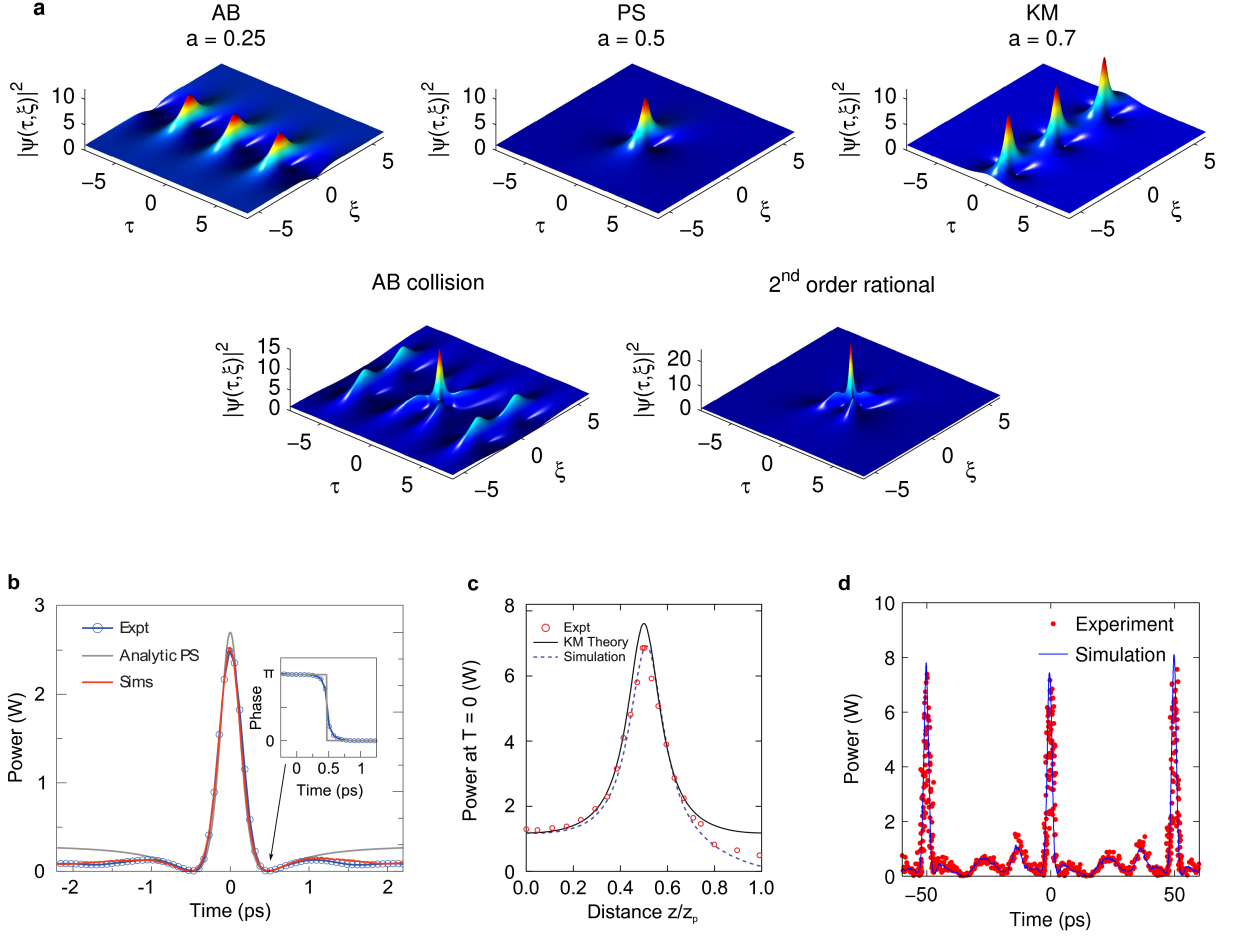

**FIGURE 1 | SFB solutions of the NLSE. a,** Analytical SFB solutions of Eq. (2) for varying parameter *a*. From left to right: an Akhmediev breather (AB); Peregrine soliton (PS); Kuznetsov-Ma (KM) soliton [42]. An example of an AB collision and the second-order rational soliton (or second-order PS) are also shown. **b-d,** Experimental results. **b** shows temporal PS properties asymptotically approached for $a = 0.42$ [39], **c** shows KM dynamics along the propagation direction for $a = 1$ [42] with experiments, simulations and theory compared in each case. Here $z_p = 5.3$ km corresponds to one period of the KM cycle. **d** compares experiments and simulations of a second-order solution consisting of the collision of two ABs ($a = 0.14$ and $a' = 0.34$) [35].

The experiments above give insight into how appropriate initial conditions can excite a range of analytic breather or SFB structures in an NLSE system. The reason why these results are important in the study of rogue waves is that structures very similar to those described by Eq. (2) (and their higher-order extensions) also *appear in chaotic fields* when MI develops from noise [33,34,47,48]. We illustrate this in Fig. 2 using numerical simulations of the dimensionless NLSE where clear signatures of the SFB solutions can be identified (i.e. periodicity in $\tau$ or along $\xi$, $\tau$-$\xi$ localization, value of peak intensity) when MI is triggered from a broadband noise background [49]. Figure 2(a) plots a density map of the evolution as a function of distance $\xi$, showing emergence of an irregular series of temporal $2\pi$, corresponding to the reciprocal of the frequency of maximum MI gain. After the initial emergence of these localized peaks, we see more complex periodic growth and decay behavior along $\xi$.

Examining particular features of the evolution map reveals signatures of the analytic KM, AB, and PS solutions described above, as plotted in Fig. 2(b). For the region marked KM, a line profile of the evolution along $\xi$ agrees well with the analytic result expected for a KM soliton, whilst in the regions AB and PS, the temporal localization characteristics also agree well with corresponding analytic predictions. Of course, observing ideal analytic SFB structures is not expected given the random initial conditions, but it is remarkable how the analytic solutions can be mapped closely to the noise-generated structures. In this context, we remark that recent results have also considered different initial forms of small perturbations to the CW background and found similar signatures of AB structures [50].

These results confirm that SFB solutions can provide analytical insight into structures emerging from noise-seeded MI, but understanding their significance to the physics of rogue waves requires analysis of the associated statistics [33,34]. To this end, Fig. 2(c) plots the histogram of the intensities of the localized peaks in a larger computational window (around $10^6$ peaks in the $\tau$–$\xi$ plane). We note firstly that the histogram maximum corresponds to intensity $|\psi|^2 = 5.6$, close to the AB intensity at maximal gain when $a = ¼$. The fact that these structures appear more frequently than others in the chaotic regime is consistent with experiments studying the spectral characteristics of spontaneous MI [18].

Secondly, we note an exponential tail (linear on a semi-logarithmic scale) for higher-intensities, and the dashed line indicates the point in the tail $I_{RW} \sim 13$ corresponding to the rogue wave "significant intensity". It is interesting to remark here that the elementary AB and PS structures (with intensity less than $I_{PS} = |\psi_{PS}|^2 = 9$) actually have intensities below the significant intensity $I_{RW}$, suggesting caution in the description of these solutions as rogue waves [24,39]. Approximately 2.5% of the total peaks have intensity exceeding the Peregrine soliton limit $I_{PS} = 9$, and these events correspond to AB collisions arising physically from the continuous range of frequencies amplified by MI [33,34]. In our simulations, the largest among them (which make up only 0.1% of the total) do have intensities exceeding the rogue wave threshold $I_{RW}$ and would very clearly be described as rogue wave events according to any criteria used [2,51]. Note that the figures below the histogram in Fig. 2(c) show simulation results comparing a typical AB solution with a third-order breather collision in the tail. Finally, we note that these results were obtained with a one-photon per mode noise background, and the exact fraction of highest-intensity events in the distribution tails is expected to vary with the noise level used to seed MI [33,34].

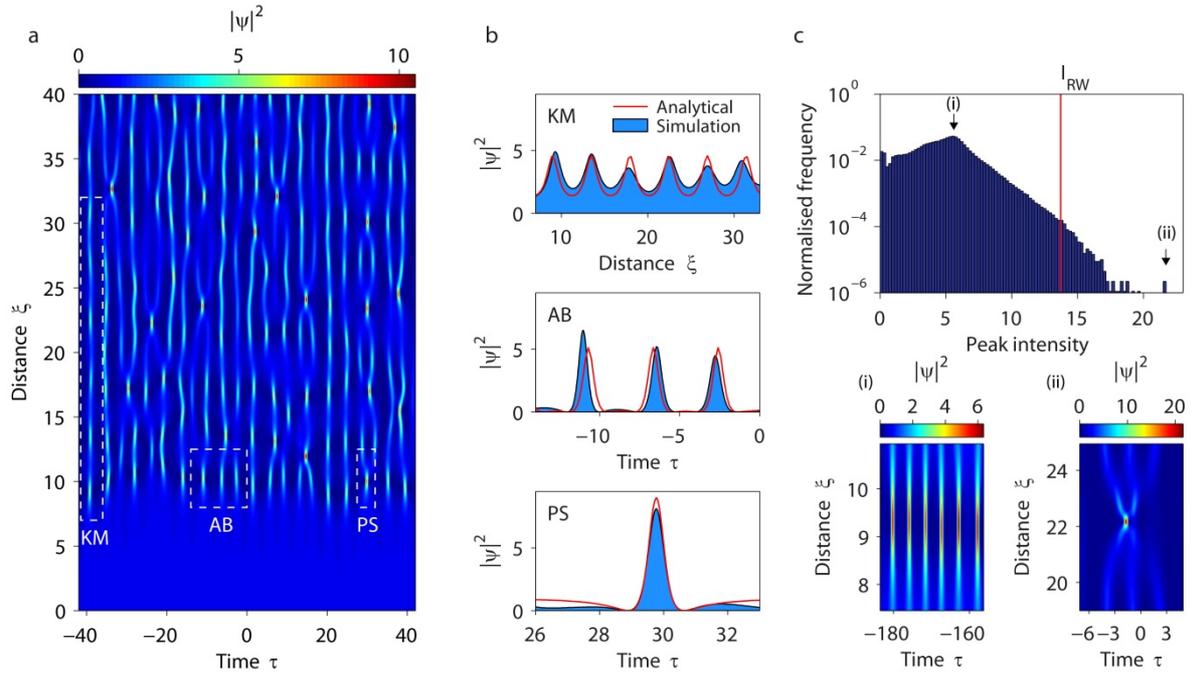

**FIGURE 2 | Numerical simulation showing signatures of analytic NLSE solutions in chaotic MI. a,** Density map showing the long term temporal evolution of a chaotic field triggered by one photon per mode noise superimposed on a CW background. Signatures of SFB solutions can be observed in the dynamics as indicated. **b,** Line profiles extracted from the regions of the chaotic field indicated by white dashed rectangles in **a,** compared with analytical NLSE solutions (red solid line). **c,** Peak-intensity statistics using an 8-connected neighbourhood regional maximum search to identify 2-D peaks from a wider simulation window. The maximum of the probability density corresponds to AB-like solutions at the peak of the MI gain, whilst the most extreme outliers arise from higher-order ABs superposition; the bottom subfigures show the evolution of two events from different regions of the histogram (i) and (ii) as indicated.

**SUPERCONTINUUM GENERATION AND SOLITONS**

The studies above show how SFB dynamics during fibre propagation can lead to strongly-localised structures with rogue wave statistics. Interestingly, however, the first observations of rogue waves in optics were not made in the regime of SFB dynamics at all, but in "long-pulse" supercontinuum generation. In this regime, higher-order effects beyond the basic NLSE played an important role [3], and although approximate SFB solutions can exist even under these conditions [52], the first observed rogue wave characteristics arose instead from the dynamics of background-free hyperbolic secant solitons.

Noise-seeded MI dynamics dominate the initial stages of long-pulse and continuous wave SC generation, but the dynamics are significantly modified with propagation by higher-order dispersion and stimulated Raman scattering [32,49]. It is this perturbed dynamical evolution that drives the emergence of the rogue wave soliton characteristics of SC generation. Specifically, after initial evolution that is statistically dominated by the AB with maximum gain (see Fig. 2), perturbations cause the associated temporal peaks to reshape with further propagation into fundamental background-free hyperbolic secant solitons [18,53]. These sech-solitons then experience additional dynamics of dispersive wave generation [54] and a continuous shift to longer wavelengths through the Raman effect [55]. Because the solitons emerge from a stochastic field of breather-like structures, their durations, amplitudes and wavelengths show considerable statistical variation. The strong dependence of the Raman self-frequency shift on duration [55], coupled with effects of group velocity dispersion (GVD), therefore induce complex subsequent evolution involving multiple (stochastic) collisions and energy exchange between the solitons [56-59].

The fact that these chaotic soliton dynamics could cause significant shot-to-shot noise in the SC spectrum has been known for many years [56], but it was the real-time measurement of these fluctuations in Ref. [3] that first highlighted links to the field of extreme events. The experiments in Ref. [3] used a long-pass wavelength filter to select out the portion of the SC spectrum where strong soliton intensity variations were expected to occur, and then used dispersive Fourier transformation [60,61] to perform shot-to-shot measurements of the associated spectral fluctuations. It was the peaks in the resulting intensity time-series that showed striking long-tailed statistics. Figure 3(a) shows the experimental setup used, and a selection of recorded histograms at three different pump levels. The main conclusion drawn from these experiments was that the largest events in the tails of the histograms corresponded to a small number of "rogue solitons" (RS) which had experienced extremely large Raman frequency shift such that their central wavelength was shifted completely within the filter transmission band (i.e. they appeared as high intensity events in the time series). More recently it has become possible to measure the full-bandwidth shot-to-shot SC spectra [62], and these experiments [see Fig. 3(b)] have highlighted more directly the small number of extremely red-shifted RSs.

Subsequent modelling and experiments examined in detail the dynamics leading to such extreme frequency shifts, clarifying the central role of soliton collisions [8,12,63-68]. Figure 3(c) shows numerical simulations highlighting the frequency and time-domain properties of a particular rogue soliton event [69]. In the frequency domain, we see the transition from initial MI dynamics with symmetric growth of noise-driven sidebands to a regime where distinct sech-soliton structures appear in the spectrum. A rogue soliton emerges at a distance of $z \sim 9$ m, and the snapshot of the time-domain evolution plot clearly illustrates a two-soliton collision at this point. The collision is associated with significant energy exchange (mediated by stimulated Raman scattering) that yields one higher energy soliton (which experiences a much greater Raman self-frequency shift) as well as a lower amplitude residual pulse [12,63,69,70].

Note that although stimulated Raman scattering plays a central role in a fibre context, any perturbation that breaks NLSE integrability can cause a homogeneous initial state to self-organize into a large-scale, coherent rogue soliton as a result of multiple interactions with other solitons and dispersive waves [71]. Indeed, numerical studies have shown that higher-order dispersive perturbations alone can give rise to rogue solitons [59,68] provided that the incoherence in the system is not too large [48,72]. Although it may seem surprising, energy exchange between colliding solitons can occur even in this case, owing to resonant coupling between the solitons and radiation emitted during the collision [73].

Other numerical studies have investigated the statistical properties of SC rogue solitons in more detail. Several authors have considered the variation in local intensity along the propagation dimension, showing that the intensity of colliding solitons at the point of collision can in fact be much higher than that of the rogue soliton at the fibre output [12,63,67]. Figure 3(d) shows this for the case of continuous wave SC generation [67], where it is clear that the maximum intensity at certain points in the fibre is much higher than at the output. This suggests that significant differences may be observed between the statistical properties measured over the full field at all points of propagation and those measured at the fibre output. A detailed study of these differences (including a discussion of the effect of spectral filtering on the statistics) has been reported [12]. We also note recent experimental work using dispersive Fourier transformation to examine the intensity correlation properties of both MI and SC to yield further insights into the underlying dynamics [61,62,74]. Finally, we remark that although most studies in optics have focused on perturbation-induced collisions as a primary generating mechanism for extreme-frequency shifting rogue solitons, the random emergence of coherent structures has been seen in numerical simulations of a basic NLSE model during the evolution of multimode CW fields with initial random phases [75].

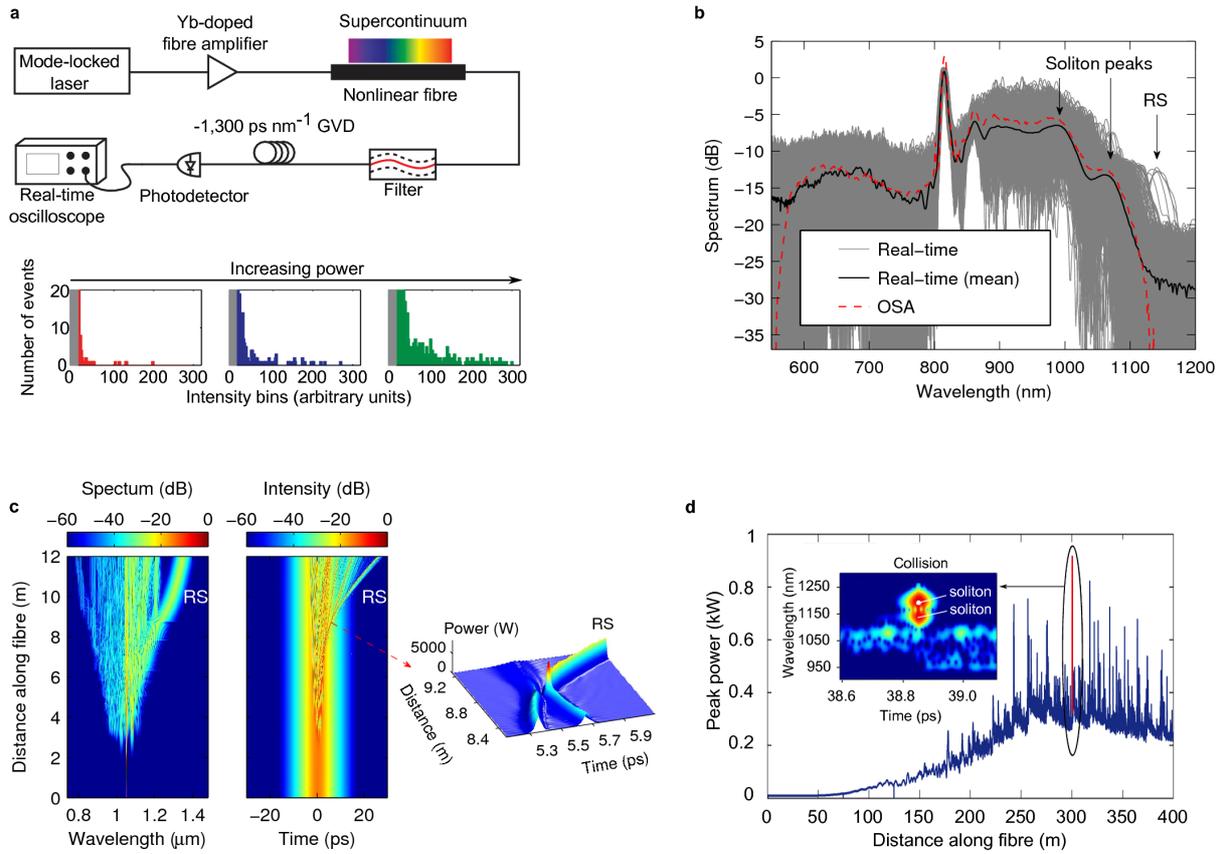

**FIGURE 3 | Selection of experimental and numerical results on supercontinuum rogue solitons. a,** Experimental setup for the first optical rogue wave soliton measurements, and the long-wavelength intensity statistics at three different pump power levels [3]. **b**, One thousand single-shot SC spectra measured using the dispersive Fourier transformation (gray); the computed mean (black); and the average spectrum measured with an optical spectrum analyser (OSA, red) [62]. **c,** Numerical simulations showing the spectral and temporal evolution of a rogue soliton in picosecond SC generation. Parameters from [69]. A rogue soliton emerges from the main spectrum at a propagation distance of 8.8 m. In the time-domain two solitons collide at the same distance, and it is the energy transfer to one of the soliton that yields the enhanced redshift. **d,** Evolution of peak intensity in numerical simulations of continuous wave SC generation. The most intense events correspond to collisions of solitons, with the time-frequency diagram shown in the inset illustrating a particular collision example [67].

**CONTROLLING ROGUE WAVES IN FIBRE SYSTEMS**

The fact that SC rogue solitons have their origin in MI suggests the possibility to control their propagation using a dual-frequency input field such that the instability develops from a coherent modulation on the input envelope rather than from noise [32]. This approach is similar to that used to excite SFBs under controlled conditions.

The potential of seeding to stabilize rogue wave dynamics was first demonstrated experimentally by controlling picosecond SC generation with a frequency-shifted replica derived from the pump pulses [76]. Although the seed pulses had only 0.01% of the pump intensity, there was dramatic improvement in the SC stability. Related numerical studies showed how an appropriate choice of seed frequency could significantly decrease the rogue soliton wavelength jitter at the same time as increasing the overall SC bandwidth [77]. These studies showed the potential of rogue soliton control in improving the performance of practical SC sources [8,77-79]. They also stimulated ideas to enhance spectral broadening in silicon waveguides [80]. Other experiments considered how seeding can improve the spectral properties of spontaneous MI, and remarkable spectral control has been demonstrated using CW seeding at only the $10^{-6}$ level [81]. A benefit of this technique is that it does not rely on the time-delay tuning required for picosecond seed pulses [82,83].

As well as controlling input conditions, other studies have shown that longitudinal variation in fibre dispersive and nonlinear properties can modify intensity fluctuations in both supercontinuum generation [84], and the dynamics of an evolving AB [85]. These studies are significant in that they show how a modified fibre propagation environment can mimic the way in which ocean topography influences water wave propagation [1].

**AMPLIFIERS AND LASERS**

In addition to the examples in conservative (or weakly dissipative) systems discussed above, rogue waves in systems with strong dissipation (gain or loss) have also been observed. Of course, dissipative dynamics in optics have been studied for decades, and resonators, amplifiers, and multimode lasers are well-known to exhibit a wide range of chaotic and self-organization effects [86]. However, the studies of optical rogue waves have now motivated the interpretation of these noise characteristics in terms of extreme value processes.

Although dissipative systems do not generally have NLSE-governed hydrodynamic counterparts, exploring regimes of long-tailed statistics in such systems has provided new insights into the underlying physical processes. The first observation of extreme events in a highly dissipative system reported long-tailed intensity statistics from Raman amplification of a coherent signal using an incoherent Raman pump [87]. The long-tailed statistics were attributed to the transfer of pump intensity fluctuations onto the coherent signal due to the exponential dependence of Raman gain on pump intensity. Similar nonlinear noise transfer underlies the emergence of extreme-value statistics in fibre parametric amplifiers [88], and silicon waveguide Raman amplifiers [89,90]. Reference [89] is notable for explicitly calculating the power-law probability distribution function of the amplified signal intensity.

The complex noise characteristics of lasers in systems with external injection, mode-locking, and delayed feedback is well known, and it is not surprising that laser noise spiking behaviour can exhibit long-tailed statistics. Experiments have reported such properties in an Erbium fibre laser with harmonic pump modulation [91], a CW Raman fibre laser [92], mode-locked Ti:Sapphire and fibre lasers [14,93,94] as well as passively-Q switched lasers [95]. In Fig. 4 we illustrate an experimental result, where a sequence of highly localized temporal noise bursts was recorded from an Erbium-doped mode-locked fibre laser; the associated intensity statistics showed significant deviation from exponentially-bounded distributions [14]. These experiments confirmed previous numerical studies predicting intensity spikes in passively mode-locked fibre lasers through chaotic pulse bunching [13,96]. Rogue wave behavior has also been seen from an optically-injected semiconductor laser, where it was shown that the rogue wave dynamics could be described deterministically, with noise influencing only the probability of their observation [97] (similar to noise-seeded MI.)

These results are inspiring significant theoretical efforts to identify the mechanisms that induce the extreme temporal localization [13,96] and to experimentally characterize the instabilities in real-time [14,94,98]. However, it should be noted that many of the reported features of extreme laser fluctuations had likely been seen in earlier experiments without being recognized as a separate class of rogue wave instability [99,100].

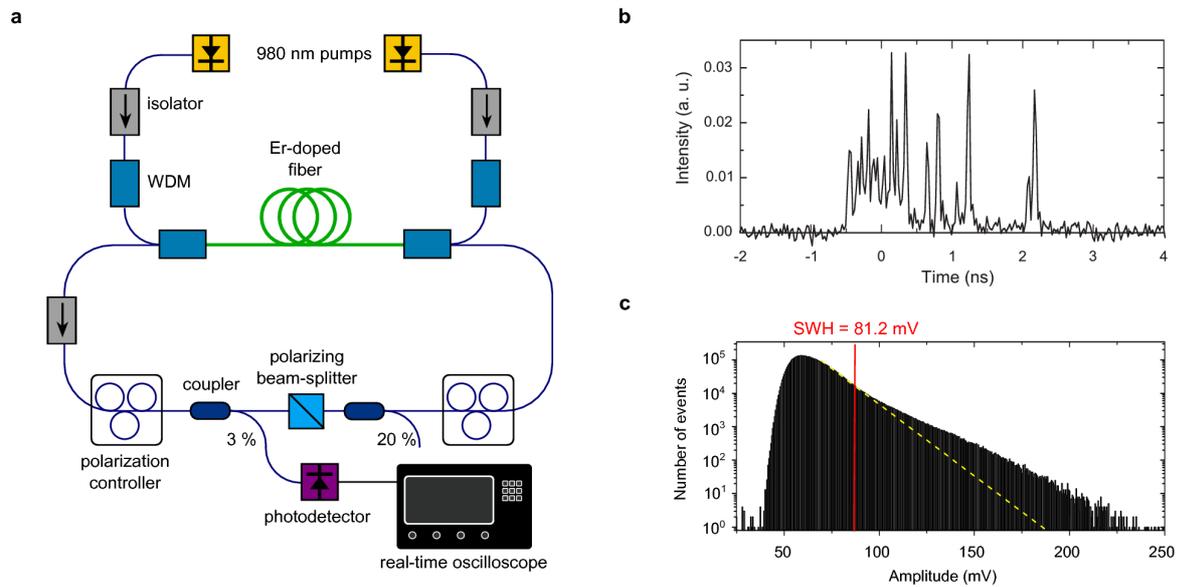

**FIGURE 4 | Dissipative rogue waves. a,** Schematic of an Er-doped mode-locked fibre laser generating a train of temporally localised bursts of noise. **b,** Time-domain experimental measurement showing an example of a chaotic pulse cluster emitted by the laser over a single roundtrip. **c,** Peak intensity statistics over nearly 10 million roundtrips. The red line shows the significant peak intensity, while the dashed yellow line corresponds to an exponentially-bounded distribution. Adapted from [14].

**SPATIAL INSTABILITIES**

There has also been extensive interest interpreting spatial instabilities in terms of extreme events. The first such study considered intensity noise in the output spatial mode of a cavity with nonlinear gain from an optically-pumped liquid-crystal light valve [101]. Depending on the system parameters, the cavity exhibited complex transverse mode dynamics which, in a high-finesse limit, showed highly unstable oscillation behavior with long-tailed statistics. The optical rogue wave-like events were attributed to the nonlinear feedback and symmetry-breaking in the cavity design.

Another example of a spatially-extended system exhibiting rogue wave statistics is optical filamentation. Filamentation is a complex process involving self-focussing, plasma formation and temporal shaping dynamics and, depending on the particular parameter regime investigated, different statistical behaviour has been reported. In the single filament regime, shot-to-shot spectral fluctuations arising from pump noise transfer through self-phase modulation have shown long-tailed statistics [9,102]. In the multi-filament regime, where the transverse profile of the input beam is broken into multiple strands through spatial MI, localized structures obeying non-Gaussian statistics have been predicted [103]. Experiments and simulations studying filamentation in gas have revealed that local refractive index variations driven by nonlinearity can induce merging of individual filament strings, giving rise to short-lived spatial rogue waves at the gas cell output [104]. Figure 5(a) shows the experimental setup in this work, with numerical and experimental results shown in Figs. 5(b) and (c), respectively.

Interestingly, long-tailed statistics have also been seen in linear spatial systems. In one experiment studying emission from a GHz microwave emitter array, it was possible to probe the emitted electromagnetic field as a function of time and spatial position across the array [105]. Although the system contained no (obvious) nonlinear element, long-tailed distributions in the microwave intensity were observed. Another example of a linear system exhibiting rogue wave statistics in the transverse spatial plane (but here at visible frequencies) was found to be the speckle pattern observed at the output of a strongly multimode fibre [106]. In this system the asymmetry and inhomogeneity in the injected beam profile [see Fig. 5(d-f)] yielded a sub-exponential intensity distribution, driving the emergence of rare high-intensity spots in the speckle pattern.

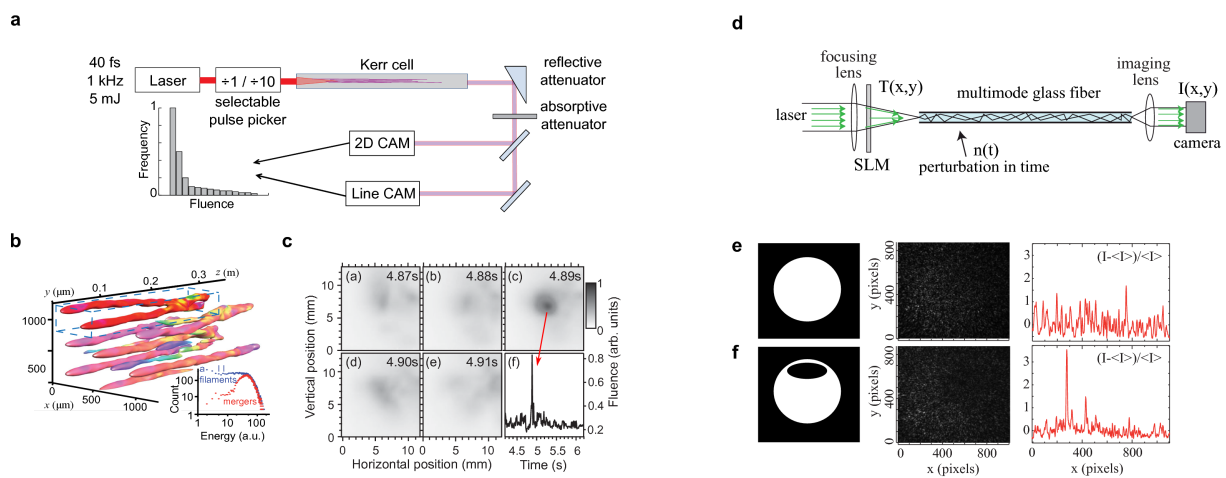

**FIGURE 5 | Rogue waves in the transverse spatial plane of optical beams. a,** Experimental setup used to study multiple filamentation in a gas cell and an example of observed fluence statistics. **b,** Results of numerical simulations, showing how highly localised rogue structures can arise from the merging of individual filaments. **c,** Two-dimensional multifilament fluence profiles experimentally measured with a high-speed camera reveals a spatio-temporal rogue wave appearing 4.89 s after the start of the recording. **d,** Experimental setup used to investigate rogue waves in the speckle pattern at the output of a multimode fibre. A spatial light modulator (SLM) is used to control the input beam profile. **e,f** Experimentally measured speckle patterns (centre) and corresponding intensity distributions taken at a selected y-coordinate (right) when the SLM transmission mask (left) is uniform (**e**) and inhomogeneous (**f**). An optical rogue wave is observed in **f**. a-c adapted from [104] and d-f adapted from [106].

**DISCUSSION AND OUTLOOK**

After the initial results of Solli *et al*. in 2007, the science of extreme events is now firmly embedded within the domain of optics, and we have seen how very different optical systems can exhibit strong localization and long-tailed statistics. However, a major conclusion of this review is that the mechanisms driving the emergence of rogue wave behaviour in optics can be very different depending on the particular system studied, and we hope that the categorisation we have provided here will assist in structuring future work in this field. We also remark that great care must be taken when comparing results obtained in different contexts. The particular example of optical fibre propagation is a case in point. The physics driving the excitation of analytical "rogue wave" breather solutions in a regime of propagation governed by the basic NLSE is significantly different from the perturbed-NLSE dynamics of extreme red-shifting rogue solitons in supercontinuum generation, and it is essential to stress this distinction.

Although there is clearly much intrinsic interest in studying extreme instabilities in optics, much of the motivation to study rogue waves in optical systems has been to gain insight into the origin of their oceanic counterparts. In this regard, however, it is also essential to recognize that not all experiments in optics directly yield insight into ocean wave propagation. There are certain regimes of wave propagation on the ocean and in optical fibre that are both well-described by a basic NLSE model and, provided experiments are performed in these regimes before the onset of any perturbations, insights obtained in the different environments can be shared. In other cases, whilst the observation of long-tailed statistics in optics may be linked to the wider theory of extreme events in physics, it is simply incorrect to compare such instabilities with oceanic wave shaping processes. Analogies can be powerful tools in physics, but they must be used with care [107].

That said, in regimes where the optical-ocean analogy is valid, there is of course intense interest to use experiments in optics to improve understanding of rogue waves in general, and experiments in optics have indeed motivated similar studies in water wave tanks (Box 1). Although rogue waves on the ocean may arise in a number of different ways, these experiments provide convincing evidence that nonlinearity in the ocean can play a role in extreme wave emergence and must be included *a priori* in any consideration of potential high amplitude ocean wave shaping mechanisms. A particular advantage of optical systems is the fact that the high repetition rate of optical sources allows the generation of large data sets such that even events occurring with extremely low probabilities can be studied [108].

Even in a strictly optical context, there remain many open directions of research. Experiments and modelling of optical rogue wave dynamics are providing new insights into how noise drives (and/or stabilises) the dynamics of nonlinear optical systems, how energy exchange occurs during soliton interactions, and how novel measurement techniques can reveal noise correlations even in the very complex supercontinuum. It is interesting to remark in closing that there is also significant effort in optics to understand how effects like dispersion and nonlinearity engineering can allow optical systems to model different aspects of the ocean environment [109]. Indeed, the recent development of "topographic" fibres have already been shown to be able to introduce control of both optical modulation instability and soliton dynamics [110,111]. As research continues to progress in this field, it is likely that the noise properties in a wide range of optical systems may find analogies with areas of physics other than oceanography, allowing the use of a convenient optical testbed with which to study a wide range of different physical processes.

**Acknowledgements**

JMD and FD acknowledge the European Research Council Advanced Grant ERC-2011-AdG-290562 MULTIWAVE. ME acknowledges the Marsden Fund of the Royal Society of New Zealand. GG acknowledges the Academy of Finland Grants 130 099 and 132 279.

**Competing Interests Statement**

The authors declare no competing financial interests.

**Correspondence**

Correspondence and requests for materials should be addressed to G.G. (email: goery.genty@tut.fi).

**BOX 1 | THE OPTICAL FIBRE-OCEAN WAVE ANALOGY**

The analogy between the dynamics of ocean waves and pulse propagation in optics arises from the central role of the NLSE in both systems. Figure B1 gives the governing equations and illustrates characteristic soliton solutions for both cases. In optics, the NLSE describes the evolution of a light pulse envelope modulating an electric field whilst for deep water, it describes the evolution of a group envelope modulating surface waves. It is important not to over-simplify or exaggerate this analogy. For example, the deep water NLSE in oceanography does not describe the shape of individual wave cycles but only their modulating envelope. Thus, specific envelope solutions of the deep water NLSE cannot be considered physically as individual "rogue waves"; within the narrowband approximation of the NLSE, there will always be multiple surface waves underneath this envelope.

The recent work studying rogue waves in optics in the MI and breather regime has motivated similar water wave experiments [112,113], even to the extent of testing the resistance of scale models of maritime vessels to particular NLSE breather solutions [114]. Interestingly, higher-order effects described by an extended NLSE can also be present in deep water [115-117], and wave tank experiments have even shown a form of hydrodynamic supercontinuum and soliton fission [118]. Note, however, that while a clear analogy between ocean wave and optical propagation exists in the unperturbed NLSE regime, there is no such rigorous analogy for the extended NLSE since the physical forms of the higher-order perturbations are very different.

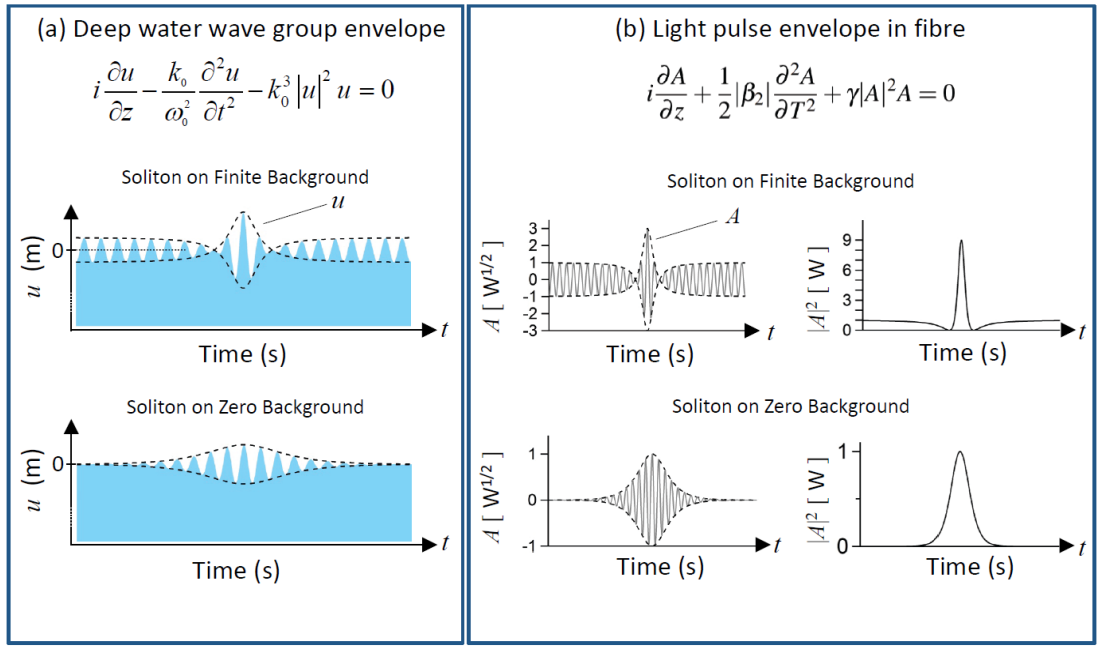

**Figure B1** | The NLSE describes evolution in a frame of reference moving at the group velocity of: (a) wave group envelopes $u$ on deep water; (b) light pulse envelopes $A$ in optical fibre with anomalous GVD. The figures illustrate solitons on finite background (top) and solitons on zero background (bottom). For the ocean wave case, there is always deep water underneath $u(z,t)$. For the water wave NLSE, $k_0$ is the wavenumber [m$^{-1}$], $\omega_0$ is the carrier frequency [rad s$^{-1}$]; for the fibre NLSE, $\beta_2 < 0$ is the GVD [ps$^2$ m$^{-1}$], $\gamma$ is the nonlinear coefficient in [W$^{-1}$ m$^{-1}$]. A water wave NLSE with time and space interchanged is also encountered but in this case the coefficients need to be adapted [2].